\documentclass[12pt]{article}
\usepackage{amssymb,amsmath,epsfig}

\begin{document}

\title{\bf Generalized Second Law of Thermodynamics in $f(T,T_{G})$ gravity}
\author{M. Zubair \thanks{mzubairkk@gmail.com; drmzubair@ciitlahore.du.pk}\\
Department of Mathematics, COMSATS Institute of
\\Information Technology Lahore, Pakistan}

\date{}

\maketitle

\begin{abstract}
An equilibrium picture of thermodynamics is discussed at the apparent horizon
of FRW universe in $f(T,T_G)$ gravity, where $T$ represents the torsion
invariant and $T_G$ is the teleparallel equivalent of the Gauss-Bonnet term.
It is found that one can translate the Friedmann equations to the standard
form of first law of thermodynamics. We discuss GSLT in the locality of
assumption that temperature of matter inside the horizon is similar to that
of horizon. Finally, we consider particular models in this theory and
generate constraints on the coupling parameter for the validity of GSLT in
terms of recent cosmic parameters and power law solutions.
\end{abstract}

\section{Introduction}

The discovery of black hole (BH) thermodynamics suggest that there
is a fundamental connection between relativistic gravity and
thermodynamics laws, however people have been trying to find a
significant way to develop such connection \cite{1}. BH acts as a
thermodynamic system with temperature being related to surface
gravity and entropy with horizon area \cite{2}. Jacobson \cite{3}
unveiled the issue of relating BH thermodynamics to the Einstein
gravity and derived Einstein field equations in local Rindler
spacetime using the entropy $S=A/4G$ and Clausius relation
$TdS=\delta{Q}$. Frolov and Kofman \cite{4} showed that for the flat
quasi de-Sitter geometry of inflationary universe, Friedmann
equations can result from $dE=TdS$ for slowly rolling scalar field.
In \cite{5}, Padmanabhan explored such connection in case of
spherically symmetric BHs and showed that field equations can be
stated in the form $dE+PdV=TdS$. This study is further extended to
generic static spacetimes in Lanczos-Lovelock gravity and shown that
the near-horizon field equations again represent a thermodynamic
identity in all these models \cite{6}.

Cai and Kim \cite{7} showed that Friedmann equations with any spatial
curvature can be derived from the Clausius relation $TdS=\delta{Q}$. The
relation between gravity and thermodynamics has also been tested in Einstein
as well as Gauss-Bonnet and Lovelock gravities. Cai and Cao \cite{8} showed
that Friedmann equations in braneworld scenario can be cast to the form of
first law of thermodynamics at the apparent horizon. This work is also
extended in the framework of warped DGP braneworld \cite{9} and Gauss-Bonnet
Braneworld \cite{10}. Akbar and Cai \cite{11} found that formulation of
thermodynamic laws in $f(R)$ and scalar tenser gravities is not trivial when
compared to Einstein gravity and Clausius relation is to be modified. In this
perspective, Eling et al. \cite{12} studied the thermodynamic laws in $f(R)$
gravity and remarked that non-equilibrium description of thermodynamics
needed, whereby the Clausius relation is modified to the form
$\delta{Q}=T(dS+d_\jmath{S})$, where $d_\jmath{S}$ is the additional entropy
term. Cai and Cao \cite{13} found that in scalar tensor theories
thermodynamics associated with the apparent horizon of the FRW universe
results in non-equilibrium description which modifies the standard Clausius
relation. The thermodynamics properties have been discussed in various
modified theories \cite{14}-\cite{20}.

The development of cosmology and gravitation can be seen as one of the
scientific triumphs of the twentieth century. In current situation modified
theories of gravity have been appeared as significant tool to discuss various
cosmic issues \cite{21}. The introduction of non-minimal coupling between
matter and curvature in the context of modified theories has become a center
of interest for the researchers \cite{22}. Another important and conceptually
rich class consists of gravitational modifications involving torsion
description of gravity. It is interesting to mention here that teleparallel
equivalent of GR has been constructed by Einstein himself by including
torsionless Levi-Civita connection instead of curvatureless Weitzenböck
connection and the vierbein as the fundamental ingredient for the theory
\cite{23}.  Harko et al. \cite{24} constructed a more general type of $f(T)$
gravity by introducing a non-minimal interaction of torsion with matter in
the Lagrangian density. We \cite{25,25a} have discussed the validity of energy
bounds and GSLT for specific models and find the feasible constraints on the involved
free parameters. Kofinas and Saridakis \cite{26} proposed a novel theory
namely $f(T_{G})$ gravity and then its generalized form $f(T,T_{G})$ gravity
and they also discussed its cosmological significance. Recently, we have
discussed the energy condition bounds in this modified gravity and tested two
well known models \cite{27} which are proposed in \cite{26}.

In this study, we are interested to explore laws of thermodynamics in
$f(T,T_{G})$ gravity which is a more generic modified theory involving
torsion and Gauss-Bonnet contributions \cite{26}. In previous studies, we
have explored the issue of equilibrium thermodynamics in $f(R,T)$ \cite{28},
$f(R,L_m)$ \cite{29} and $f(R,T,R_{\mu\nu}T^{\mu\nu})$ \cite{30} theories of
gravity. We find that equilibrium picture of thermodynamics in such theories
needs more study to follow. However, in this paper we find that one can
develop the equilibrium picture of thermodynamics in generic modified theory
models which involve contribution from torsion scalar. The paper has the
following format: In section \textbf{2}, we present the general formalism of
field equation in $f(T,T_G)$ gravity for FRW universe. In section \textbf{3},
the first law of thermodynamics (FLT) is established and we discuss the
validity of GSLT for different $f(T,T_G)$ models in section \textbf{4}.
Finally, section \textbf{5} summarizes our findings.

\section{$f(T,T_{G})$ Gravity}

In this section, we briefly review some basic components of TEGR and hence of
$f(T,T_{G})$. The dynamical variables of TEGR are the vielbein fields
$e_A(x^\mu)$ which can be represented in components as
$e_a=e_a^{\mu}\partial_{\mu}$. The connection 1-forms $\omega^a_b(x^\mu)$
(the source of parallel transportation) in terms of vielbein field is given
by $\omega^{a}_{b}=\omega^{a}_{b\mu}dx^\mu=\omega^{a}_{bc}e^{c}$. The
structure coefficients arising from the veilbein commutation relation
\begin{equation*}
[e_{a}, e_{b}]=C^{c}_{ab}e_{c},
\end{equation*}
are defined by
\begin{equation*}
C^{c}_{ab}=e_{a}^{\mu}e_{b}^{\nu}(e^{c}_{\mu,\nu}-e^{c}_{\nu,\mu}).
\end{equation*}
One can define torsion and curvature tensor in tangent components as
\begin{eqnarray}\label{1}
T^{a}_{bc}&=&\omega^{a}_{cb}-\omega^{a}_{bc}-C^{a}_{bc},\\\label{2}
R^{a}_{bcd}&=&\omega^{a}_{bd,c}-\omega^{a}_{bc,d}+\omega^{e}_{bd}\omega^{a}_{ec}
-\omega^{e}_{bc}\omega^{a}_{ed}-C^{e}_{cd}\omega^{a}_{be}.
\end{eqnarray}
Furthermore, for an orthonormal veilbein, the metric tensor is defined by the
relation
\begin{equation*}
g_{\mu\nu}=\eta_{ab}e^a_\mu{e}^b_\nu,
\end{equation*}
where $\eta_{ab}=diag(-1,1,1,1)$.
Herein, a,b run over $0,1,2,3$ for the tangent space of the manifold and
$\mu,\nu$ are coordinate indices on the manifold which also run over
$0,1,2,3$.

In order to be consistent with the condition $R^{a}_{bcd}=0$
(teleparallelism condition), we express the Weitzenb$\ddot{o}$ck
connection as follows
$$\tilde{\omega}^{\lambda}_{\mu\nu}=e_{a}^{\lambda}e^{a}_{\mu,\nu},$$
while in terms of Levi-Civita connection, the Ricci scalar $R$ is given by
\begin{eqnarray}\nonumber
e\bar{R}=-eT+2(eT_{\nu}^{\nu\mu})_{,\mu},
\end{eqnarray}
where $e=det(e^{a}_{\mu})=\sqrt{|g|}$ and
$T=\frac{1}{4}T^{\mu\nu\lambda}_{\mu\nu\lambda}
+\frac{1}{2}T^{\mu\nu\lambda}T_{\lambda\nu\mu}-T_{\nu}^{\nu\mu}T^{\lambda}_{\lambda\mu}$
is the torsion scalar. Consequently, the Lagrangian density describing TEGR
in D-dimensions is given by
\begin{eqnarray}\label{3}
S_{tel}=-\frac{1}{2\kappa_{D}^2}\int_M d^Dx eT.
\end{eqnarray}
Following these lines TEGR action has been extended to the form \cite{21}
\begin{equation}\label{4}
\mathcal{I}=\frac{1}{2\kappa^2}\int{dx^4ef(T)}.
\end{equation}
In a recent paper \cite{26}, teleparallel equivalent of Gauss-Bonnet theory
has been proposed involving a new torsion scalar $T_{G}$, where, in
Levi-Civita connection, the Gauss-Bonnet term is defined by
\begin{equation}\nonumber
e\bar{G}=eT_{G}+\verb"total diverg",
\end{equation}
and the corresponding action takes the following form
\begin{eqnarray}\label{5}
S_{tel}=-\frac{1}{2\kappa_{D}^2}\int_M d^Dx eT_{G}.
\end{eqnarray}
Since both theories $f(T)$ and $f(T_G)$ arise independently, therefore a
comprehensive theory involving both $T$ and $T_G$ as basic ingredient has
been proposed by Kofinas and Saridakis defined by the following action
\cite{26}
\begin{eqnarray}\label{6}
S=-\frac{1}{2\kappa_{D}^2}\int_M d^Dx ef(T,T_{G}).
\end{eqnarray}
In some certain limits of the function $f(T,T_{G})$, other theories like GR,
TEGR, Einstein-Gauss-Bonnet theory etc. can be discussed.

We consider the flat FRW universe model with $a(t)$ as expansion scalar given
by
\begin{eqnarray}\label{7}
ds^2=-dt^2-a^2(t)(dx^2+dy^2+dz^2).
\end{eqnarray}
The diagonal vierbein and the dual vierbein for this metric are
\begin{eqnarray}\nonumber
e^a_\mu=diag(1, a(t), a(t), a(t)),\\\nonumber e^\mu_a=(1, a^{-1}(t),
a^{-1}(t), a^{-1}(t)),
\end{eqnarray}
while the corresponding determinant is given by $e=a(t)^3$. The torsion
scalar and Gauss-Bonnet equivalent term $T_G$ for this geometry are
\begin{equation}\label{8}
T=6H^2, \quad T_G=24H^2(\dot{H}+H^2),
\end{equation}
where $H=\frac{\dot{a}}{a}$ is the Hubble parameter. In this study, we
consider the matter action  $S_m=\int\mathcal{L}_{matter}\sqrt{-g}dx^4$
corresponding to natter energy momentum tensor $\Theta_{\mu\nu}$, which is
assumed as perfect fluid. Now the variation of action $S+S_m$ implies the
following gravitational equations for FRW geometry
\begin{eqnarray}\label{9}
&&f-12H^2f_{T}-T_{G}f_{T_{G}}+24H^3\dot{f}_{T_{G}}=2\kappa^2\rho_m,\\\nonumber
&&f-4(\dot{H}+3H^2)f_T-4H\dot{f}_T-T_Gf_{T_{G}}+\frac{2}{3H}T_G\dot{f}_{T_{G}}
+8H^2\ddot{f}_{T_{G}}=-2\kappa^2p_m,\\\label{10}
\end{eqnarray}
where $\rho_m$ and $p_m$ indicates the density and pressure of ordinary
matter, $f_{TT},~f_{TT_G},...$ represent the second and higher-order
derivatives with respect to $T$ and $T_G$ respectively. Moreover dot
represents the time derivative and these derivatives are given by
\begin{eqnarray}\label{11}
\dot{f}_T&=&f_TT\dot{T}+f_{TT_G}\dot{T}_G,\quad
\dot{f}_{T_G}=f_{TT_G}\dot{T}+f_{T_GT_G}\dot{T}_G, \\\label{12}
\ddot{f}_{T_G}&=&f_{TTT_G}\dot{T}^2+2f_{TT_GT_G}\dot{T}\dot{T}_G+f_{T_GT_GT_G}\ddot{T}_G^2
+f_{TT_G}\ddot{T}+f_{T_GT_G}\ddot{T}_G,
\end{eqnarray}
where the derivatives of torsion scalar $T$ and teleparallel equivalent to
Gauss-Bonnet term $T_G$ can be set in terms of Hubble parameter $H$ as
\begin{eqnarray}\nonumber
\dot{T}&=&12H\dot{H}, \quad
\dot{T}_G=24H^2(\ddot{H}+2H\dot{H})+48H\dot{H}(\dot{H}+H^2),\\\nonumber
\ddot{T}&=&12(\dot{H}^2+H\ddot{H}), \quad
\ddot{T}_G=48\dot{H}^3+144H\dot{H}\ddot{H}+288\dot{H}^2H^2+24H^2\dddot{H}\\\nonumber
&+&96H^3\ddot{H}.
\end{eqnarray}
The dynamical equations (\ref{9}) and (\ref{10}) can be rewritten as
\begin{eqnarray}\label{13}
H^2&=&\frac{\kappa^2}{3}(\rho_m+\rho_{\vartheta}),\\\label{14}
\dot{H}&=&-\frac{\kappa^2}{2}(\rho_m+p_m+\rho_{\vartheta}+p_{\vartheta}),
\end{eqnarray}
where $\rho_{\vartheta}$ and $p_{\vartheta}$ are the density and pressure of
dark energy, respectively given by
\begin{eqnarray}\label{15}
\rho_{\vartheta}&=&\frac{1}{2\kappa^2}[6H^2-f+12H^2f_T+T_Gf_{T_{G}}-24H^3\dot{f}_{T_G}],\\\nonumber
p_{\vartheta}&=&\frac{1}{2\kappa^2}[-2(2\dot{H}+3H^2)+f-4(\dot{H}+3H^2)f_T-4H\dot{f}_T-T_Gf_{T_G}\\\label{16}
&+&\frac{2}{3H}T_G\dot{f}_{T_G}+8H^2\ddot{f}_{T_G}].
\end{eqnarray}
For FRW spacetime, the energy density $\rho_{\vartheta}$ and pressure
$p_{\vartheta}$ of torsion contributions satisfy the following relation
\begin{equation}\label{17}
\dot{\rho_\vartheta}+3H(\rho_\vartheta+p_\vartheta)=0.
\end{equation}

\section{Thermodynamics in $f(T,T_G)$ Gravity}

Here, we discuss the first and second laws of thermodynamics at the apparent
horizon of FRW universe in $f(T,T_G)$ gravity.

\subsection{First Law of Thermodynamics}

The condition
$h^{\alpha\beta}\partial_{\alpha}\tilde{r}\partial_{\beta}\tilde{r}=0$,
implies the radius of dynamical apparent horizon. For flat FRW geometry,
radius $\tilde{r}_A$ is
\begin{equation}\label{18}
\tilde{r}_A=\frac{1}{H},
\end{equation}
 Taking the time derivative of the above equation,
it follows that
\begin{equation}\nonumber
\frac{\tilde{r}^{-3}_A d\tilde{r}_A}{dt}=H\dot{H}.
\end{equation}
Substituting above result in Eq.(\ref{14}), we get
\begin{eqnarray}\nonumber
&&\frac{1}{2\pi\tilde{r}_A}\left(\frac{2\pi\tilde{r}_Ad\tilde{r}_A}{G}\right)=
4\pi\tilde{r}_A^3H\left[\rho_m+p_m+\frac{1}{16\pi{G}}\left(-4\dot{H}(1+f_T)-4H\dot{f_T}
\right.\right.\\\label{19}&+&\left.\left.(16H\dot{H}-8H^3)\dot{f_{T_G}}+8H^2\ddot{f_{T_G}}\right)\right]dt.
\end{eqnarray}
Multiplying Eq.(\ref{19}) with
$\left(1-\frac{\dot{\tilde{r}}_A}{2H\tilde{r}_A}\right)$, it leads to
\begin{eqnarray}\nonumber
&&\frac{|\kappa_{sg}|}{2\pi}d\left(\frac{A}{4G}\right)=\left(1-\frac{\dot{\tilde{r}}_A}{2H\tilde{r}_A}\right)
\tilde{r}_A{A}H\left[\rho_m+p_m+\frac{1}{16\pi{G}}\left(-4\dot{H}(1+f_T)\right.\right.\\\label{20}&-&\left.\left.4H\dot{f_T}
+(16H\dot{H}-8H^3)\dot{f_{T_G}}+8H^2\ddot{f_{T_G}}\right)\right]dt,
\end{eqnarray}
where $A=4\pi{\tilde{r}^2_A}$ is the area of apparent horizon,
$\kappa_{sg}=\frac{1}{\tilde{r}_A}(1-\frac{\dot{\tilde{r}}_A}{2H\tilde{r}_A})$
is the surface gravity and $T=\frac{|\kappa_{sg}|}{2\pi}$ is identified as
temperature of apparent horizon. Furthermore, Eq.(\ref{20}) involves
Bekenstein-Hawking entropy relation \cite{1,2} $S=A/4G$ defined in Einstein
gravity. Consequently, Eq.(\ref{20}) can be rewritten as
\begin{eqnarray}\nonumber
&&TdS=\tilde{r}_A{A}H(\rho_m+p_m)dt-2\pi\tilde{r}_A^2(\rho_m+p_m)d\tilde{r}_A
+\frac{(\tilde{r}_A{A}H-2\pi\tilde{r}_A^2\dot{\tilde{r}}_A)}{16\pi{G}}\\\label{21}&\times&\left(-4\dot{H}(1+f_T)
-4H\dot{f_T}+(16H\dot{H}-8H^3)\dot{f_{T_G}}+8H^2\ddot{f_{T_G}}\right)dt.
\end{eqnarray}
The matter energy density inside the apparent horizon is defined by the
relation $E={V}\rho_{tot}$ with $V=4/3\pi\tilde{r}_A^3$ and for this theory
it results in
\begin{equation}\label{22}
dE=4\pi{\tilde{r}}_A^2(\rho_m+\rho_\vartheta){d}\tilde{r}_A-4\pi\tilde{r}_A^3(\rho_m+\rho_\vartheta+p_m+p_{\vartheta})Hdt,
\end{equation}
where we have employed the standard continuity equation which holds in
$f(T,T_G)$ gravity.

Inserting $dE$ in Eq.(\ref{21}), we get
\begin{equation}\label{23}
TdS=-dE+2\pi\tilde{r}_A^2(\rho_m+\rho_\vartheta-p_m-p_{\vartheta})d\tilde{r}_A.
\end{equation}
Now introducing the total work density which is defined as \cite{32}
\begin{equation}\label{24}
W=-\frac{1}{2}T^{(tot)\alpha\beta}h_{\alpha\beta}=\frac{1}{2}
(\rho_{tot}-p_{tot}).
\end{equation}
Eq.(\ref{23}) takes the standard form of FLT
\begin{equation}\label{25}
TdS=-dE+dW,
\end{equation}
which is FLT in $f(T,T_G)$ gravity identical to that in Einstein,
Gauss-Bonnet and Lovelock gravities and usual FLT is satisfied by the
respective field equations \cite{8}-\cite{10}.

\section{GSLT in $f(T,T_{G})$ Gravity}

Here, we explore the validity of GSLT in the framework of $f(T,T_{G})$
gravity at the apparent horizon. According to GSLT, the sum of the horizon
entropy and entropy of ordinary matter fluid components is not decreasing
with time \cite{7}. In literature, it is shown that GSLT can be met in the
framework of modified theories of gravity \cite{14}-\cite{20,28,29,30}. It
would be interesting to examine the GSLT in $f(T,T_{G})$ modified theory. The
Gibb's equation which relates the entropy of matter and energy sources inside
the horizon $S_{in}$ to the density and pressure in the horizon is defined as
\begin{equation}\label{26}
T_{in}dS_{in}=d(\rho_mV)+p_{m}dV.
\end{equation}
Equivalently, it can be expressed as
\begin{equation}\label{27}
T_{in}\dot{S}_{in}=4\pi{\tilde{r}^2_A}(\rho_m+p_m)(\dot{\tilde{r}}_A-H\tilde{r}_A),
\end{equation}
where $\rho_m$ and $p_m$ can be evaluated of the form
\begin{eqnarray}\label{28}
\rho_m&=&\frac{1}{2}\left[f-12H^2f_{T}-T_{G}f_{T_{G}}+24H^3\dot{f}_{T_{G}}\right],\\\nonumber
p_m&=&\frac{1}{2}\left[-f+4(\dot{H}+3H^2)f_T+4H\dot{f}_T+T_Gf_{T_{G}}-\frac{2}{3H}T_G\dot{f}_{T_{G}}
-8H^2\ddot{f}_{T_{G}}\right].\\\label{29}
\end{eqnarray}
Substituting Eqs.(\ref{27}) and (\ref{28}) in Eq.(\ref{27}), it follows
\begin{eqnarray}\label{30}
T_{in}\dot{S}_{in}=2\pi{\tilde{r}^2_A}(\dot{\tilde{r}}_A-4H\tilde{r}_A)\left[\partial_t(4Hf_T)
+(8H^3-16H\dot{H})\dot{f}_{T_{G}}-8H^2\ddot{f}_{T_{G}}\right].
\end{eqnarray}
Using the horizon entropy relation, one can find
\begin{equation}\label{31}
T_h\dot{S}_h=\frac{1}{2\tilde{r}_AHG}(2H\tilde{r}_A\dot{\tilde{r}}_A-\dot{\tilde{r}}^2_A).
\end{equation}
Now we discuss the validity of GSLT which requires
$(\dot{T}_h\dot{S}_h+\dot{T}_{in}\dot{S}_{in})\geqslant0$. In this setting,
we assume a relation between the temperature of matter and energy sources
within the horizon and temperature of apparent horizon \emph{i.e.},
$T_{in}=bT_h$, where $0<b<1$. It is natural to assume a relation between the
temperature of apparent horizon and entire contents within the horizon which
results in thermal equilibrium for the choice of $b=1$. Generally speaking,
the horizon temperature varies from the temperature of all energy sources
inside the horizon and this variation makes the spontaneous flow of energy
between between the horizon and fluid components so that thermal equilibrium
is no longer preserved \cite{32}. Here, we are discussing the equilibrium
description of thermodynamics in $f(T,T_G)$ gravity, so that we limit our
results to the case of thermal equilibrium $b=1$ \emph{i.e.}, the horizon
temperature is equal to that of fluid components inside the horizon.

After some manipulation Eqs.(\ref{30}) and (\ref{31}) can be summed to the
following form
\begin{eqnarray}\nonumber
T_h\dot{S}_{tot}&=&\frac{-\dot{H}}{2GH^4}(2H^2+\dot{H})-\frac{2\pi}{H^4}(H^2+\dot{H})\left[
\partial_t(4Hf_T)+(8H^3-16H\dot{H})\dot{f}_{T_{G}}\right.\\\label{32}&-&\left.8H^2\ddot{f}_{T_{G}}\right],
\end{eqnarray}
which is a condition to validate the GSLT in $f(T,T_G)$ gravity and it can be
varified for different choices of Lagrangian. To illustrate the validity of
GSLT in $f(T,T_G)$ gravity, we consider some generic $f(T,T_G)$ models of the
following form \cite{26}
\begin{enumerate}
\item $f(T,T_G)=-T+\alpha_1\sqrt{T^2+\alpha_2{T_G}}$,
\item
    $f(T,T_G)=-T+\beta_1\sqrt{T^2+\beta_2T_G}+\alpha_1T^2+\alpha_2T\sqrt{|T_G|}$,
\item $f(T,T_G)=-T+\beta_1(T^2+\beta_2T_G)+\beta_3(T^2+\beta_4T_G)^2$,
\end{enumerate}
where $T_G$ contains the quartic torsion term, $\alpha_i$'s and $\beta_i$'s
are dimensionless coupling parameters. These models have been proposed in
\cite{26}, where authors discussed the phase space analysis and expansion
history from early-times inflation to late-times cosmic acceleration with no
need of introducing cosmological constant. It is found that effective EoS
parameter can represents different eras of the universe namely, quintessence,
phantom and quintom phase (crossing of phantom divide line).
\begin{itemize}
  \item $f(T,T_G)=-T+\alpha_1\sqrt{T^2+\alpha_2{T_G}}$
\end{itemize}
Initially, we consider the model of the form
$f(T,T_G)=-T+\alpha_1\sqrt{T^2+\alpha_2{T_G}}$, where $\alpha_i$'s are
constrained for the validity of GSLT. The derivatives of $f$ can be
calculated as
\begin{eqnarray}\nonumber
f_T&=&-1+\alpha_1T(T^2+\alpha_2{T_G})^{-1/2},\\\nonumber
\dot{f}_T&=&\alpha_1\dot{T}(T^2+\alpha_2{T_G})^{-1/2}-\frac{\alpha_1}{2}(2T^2\dot{T}+\alpha_2T\dot{T}_G)(T^2+\alpha_2{T_G})^{-3/2},\\\nonumber
\dot{f}_{T_G}&=&\frac{-\alpha_1\alpha_2}{4}(T^2+\alpha_2{T_G})^{-3/2}(2T\dot{T}+\alpha_2\dot{T}_G),\\\nonumber
\ddot{f}_{T_G}&=&\frac{\alpha_1\alpha_2}{4}(T^2+\alpha_2{T_G})^{-3/2}\left[\frac{3}{2}(T^2+\alpha_2{T_G})^{-1}(2T\dot{T}+\alpha_2\dot{T}_G)
+(2\dot{T}^2\right.\\\nonumber&+&\left.2T\ddot{T}+\alpha_2\ddot{T}_G)\right].
\end{eqnarray}
Using the above relations, one can find GSLT of the following form
\begin{eqnarray}\nonumber
T_h\dot{S}_{tot}&=&\frac{-\dot{H}}{2GH^4}(2H^2+\dot{H})-\frac{2\pi}{H^4}(H^2+\dot{H})\left[4\dot{H}(-T
+\alpha_1\sqrt{T^2+\alpha_2{T_G}})\right.\\\nonumber&+&\left.4H\{\alpha_1\dot{T}(T^2+\alpha_2{T_G})^{-1/2}-\frac{\alpha_1}{2}(2T^2\dot{T}
+\alpha_2T\dot{T}_G)(T^2+\alpha_2{T_G})^{-3/2}\}\right.\\\nonumber&+&\left.(8H^3-16H\dot{H})\{\frac{-\alpha_1\alpha_2}{4}(T^2+\alpha_2{T_G})^{-3/2}
(2T\dot{T}+\alpha_2\dot{T}_G)\}-8H^2\right.\\\nonumber&\times&\left.\{\frac{\alpha_1\alpha_2}{4}(T^2+\alpha_2{T_G})^{-3/2}\{\frac{3}{2}(T^2
+\alpha_2{T_G})^{-1}(2T\dot{T}+\alpha_2\dot{T}_G)+(2\dot{T}^2\right.\\\label{33}&+&\left.2T\ddot{T}+\alpha_2\ddot{T}_G)\}\}\right].
\end{eqnarray}
Here, we define some cosmic parameters namely deceleration, jerk and snap
parameters in terms of $H$ as
\begin{equation}\nonumber
q=-\left(1+\frac{\dot{H}}{H^2}\right),\quad r=2q^2+q-\frac{\dot{q}}{H},\quad
s=\frac{(r-1)}{3(q-1/2)},
\end{equation}
so that the time derivatives of $H$ can be expressed in terms of these
parameters as
\begin{eqnarray}\nonumber
\dot{H}&=&-H^2(1+q),\quad \ddot{H}=H^3(j+3q+2),\quad
\dddot{H}=H^4(s-4j-3q(q+4)-6).
\end{eqnarray}
Hence, one can represent $T,~T_G$ and their derivatives in terms of recent
value of Hubble parameter $H_0$ and cosmic parameters of the following form
\begin{eqnarray}\nonumber
T&=&6H_0^2, \quad T_G=-24qH_0^4, \quad
\dot{T}=12H_0^3(1+q),\\\nonumber
\dot{T}_G&=&48H_0^5(1+q)^2-96H_0^5(1+q)+24H_0^5(j+3q+2),\\\nonumber
\ddot{T}&=&12H_0^4(1+q)^2+12H_0^4(j+3q+2),\\\nonumber
\ddot{T}_G&=&-48H_0^6(1+q)^3-144H_0^6(1+q)(j+3q+2)+288H_0^6(1+q)^2\\\label{34}
&+&24H_0^6(s-4j-3q(q+4)-6)+96H_0^6(j+3q+2).
\end{eqnarray}
Substituting relations (\ref{34}) in (\ref{33}), it implies the GSLT in terms
of recent values of cosmic parameters. In this study, we set the present day
values of Hubble, deceleration, jerk and snap parameters as
$H_0=73.8,~q_0=-0.81\pm0.14,~j_0=2.16^{+0.81}_{-0.75}$ and
$s_0=-0.22^{+0.21}_{-0.19}$ \cite{32}. In Figure 1, we show the evolution of
GSLT for model 1 in terms of parameters $\alpha_1$ and $\alpha_2$. It can be
seen that GSLT is satisfied for $\alpha_i>0$.
\begin{figure}
\centering \epsfig{file=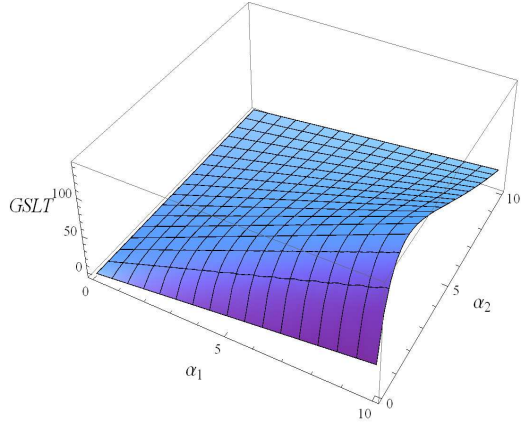, width=.495\linewidth,
height=2.2in} \caption{Evolution of GSLT for the model
$f(T,T_G)=-T+\alpha_1\sqrt{T^2+\alpha_2{T_G}}$ in terms of coupling parameters $\alpha_1$
and $\alpha_2$.}
\end{figure}

Cosmic expansion history is thought to have experienced the decelerated phase
and hence transition to accelerating epoch. Thus, power law solutions can
play vital role to connect the matter dominated phase with accelerating
paradigm. The existence of power law solutions in FRW setting is particularly
relevant to intimate all possible cosmic evolutions. The scale factor for
power law cosmology is defined as
\begin{equation}\nonumber
a(t)=a_0t^m,
\end{equation}
where $m$ is a positive real number. If $0<m<1$, then the required power law
solution is decelerating while for $m>1$ it exhibits accelerating behavior.
To be more explicit for the above constraint, we set the power law cosmology
for accelerated cosmic expansion ($m>1$). For FRW universe, we show the
evolution of GSLT in Figure 2. It can be seen that validity of GSLT requires
$m>2$ with $\alpha_1=14$ and $\alpha_2=0.02$.
\begin{figure}
\centering \epsfig{file=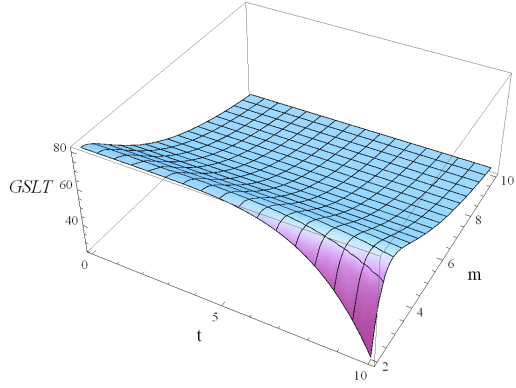, width=.49\linewidth,
height=2in} \caption{Evolution of GSLT for the model
$f(T,T_G)=-T+\alpha_1\sqrt{T^2+\alpha_2{T_G}}$ versus $m$ and $t$ with coupling parameters $\alpha_1=14$
and $\alpha_2=0.02$.}
\end{figure}
\begin{itemize}
  \item
      $f(T,T_G)=-T+\beta_1\sqrt{T^2+\beta_2T_G}+\alpha_1T^2+\alpha_2T\sqrt{|T_G|}$
\end{itemize}
This model is modified version of previous model and involves higher order
correction terms like $T^2$ and $T\sqrt{|T_G|}$. Here, one can find the
derivatives $f$ as
\begin{eqnarray}\nonumber
\dot{f}_T&=&-1+\beta_1T\dot{T}(T^2+\beta_2T_G)^{-1/2}+2\alpha_1T\dot{T}+\alpha_2\dot{T}\sqrt{|T_G|},\\\nonumber
\dot{f}_T&=&[\{\beta_1(T^2+\beta_2T_G)^{-1/2}+2\alpha_1\}\dot{T}-\{\beta_1T^2\dot{T}
+\frac{1}{2}\beta_1\beta_2T\dot{T}_G\}(T^2+\beta_2T_G)^{-3/2}\\\nonumber
&+&\frac{\alpha_2}{2}(|T_G|)^{-1/2}\dot{T}_G],\\\nonumber
\dot{f}_{T_G}&=&[\{-\frac{1}{2}\beta_1\beta_2T(T^2+\beta_2T_G)^{-3/2}
+\frac{\alpha_2}{2}(|T_G|)^{-1/2}\}\dot{T}+\{-\frac{\beta_1\beta_2^2}{4}(T^2+\beta_2T_G)^{-3/2}\\\nonumber
&-&\frac{\alpha_2T}{4}(|T_G|)^{-3/2}\}\dot{T}_G],\\\nonumber
\ddot{f}_{T_G}&=&\{-\frac{\beta_1\beta_2}{2}(T^2+\beta_2T_G)^{-3/2}
+\frac{3\beta_1\beta_2}{2}T^2(T^2+\beta_2T_G)^{-5/2}\}\dot{T}^2+2\{\frac{3}{4}\beta_1\beta_2^2T(T^2\\\nonumber
&+&\beta_2T_G)^{-5/2}-\frac{\alpha_2}{4}(|T_G|)^{-3/2}\}\dot{T}\dot{T}_G
+\{\frac{3}{8}\beta_1\beta_2^3(T^2+\beta_2T_G)^{-5/2}
\\\label{35}&+&\frac{3\alpha_2T}{8}(|T_G|)^{-5/2}\}\ddot{T}_G.
\end{eqnarray}
Using the derivatives (\ref{35}), we can represent the GSLT as
\begin{eqnarray}\nonumber
T_h\dot{S}_{tot}&=&\frac{-\dot{H}}{2GH^4}(2H^2+\dot{H})-\frac{2\pi}{H^4}(H^2+\dot{H})
\left[4H\{(\beta_1(T^2+\beta_2T_G)^{-1/2}+2\alpha_1)\dot{T}\right.\\\nonumber&-&\left.(\beta_1T^2\dot{T}
+\frac{1}{2}\beta_1\beta_2T\dot{T}_G)(T^2+\beta_2T_G)^{-3/2}
+\frac{\alpha_1}{2}(|T_G|)^{-1/2}\dot{T}_G\}+4\dot{H}\{-1\right.\\\nonumber&+&\left.\beta_1T\dot{T}(T^2
+\beta_2T_G)^{-1/2}+2\alpha_1T\dot{T}+\alpha_2\dot{T}\sqrt{|T_G|}\}+(8H^3-16H\dot{H})
\right.\\\nonumber&\times&\left.\{(\frac{-\beta_1\beta_2}{2}T(T^2+\beta_2T_G)^{-3/2}+\frac{\alpha_2}{2}(|T_G|)^{-1/2})\dot{T}
+(-\frac{\beta_1\beta_2^2}{4}(T^2+\beta_2T_G)^{-3/2})\right.\\\nonumber&-&\left.\frac{\alpha_2T}{4}(|T_G|)^{-3/2})\dot{T}_G\}
-8H^2\{(-\frac{\beta_1\beta_2}{4}(T^2+\beta_2T_G)^{-3/2}+\frac{3\beta_1\beta_2}{2}T^2(T^2\right.\\\nonumber&+&\left.\beta_2T_G)^{-5/2})
\dot{T}^2+(\frac{3}{2}\beta_1\beta_2^2T(T^2+\beta_2T_G)^{-5/2}-\frac{\alpha_2}{2}(|T_G|)^{-3/2})\dot{T}\dot{T}_G
\right.\\\label{36}&+&\left.(\frac{3}{8}\beta_1\beta_2^3(|T_G|)^{-5/2}+\frac{3\alpha_2T}{8}(|T_G|)^{-5/2})\ddot{T}_G\}\right].
\end{eqnarray}
We first analyze the evolution of GSLT in terms of present day values of
cosmic parameters and show the respective behavior in Figure 3. In left plot
we present the validity of GSLT in terms of parameters $\beta_i$, which can
be met only if $\alpha_i<0$. Similarly, in right plot we fix $\beta_i$ and
GSLT is satisfied if $\alpha_i<0$. Furthermore, we consider the power law
cosmology and present the validity of GSLT in term of $t$ and $m$ as shown in
Figure 4.
\begin{figure}
\center\epsfig{file=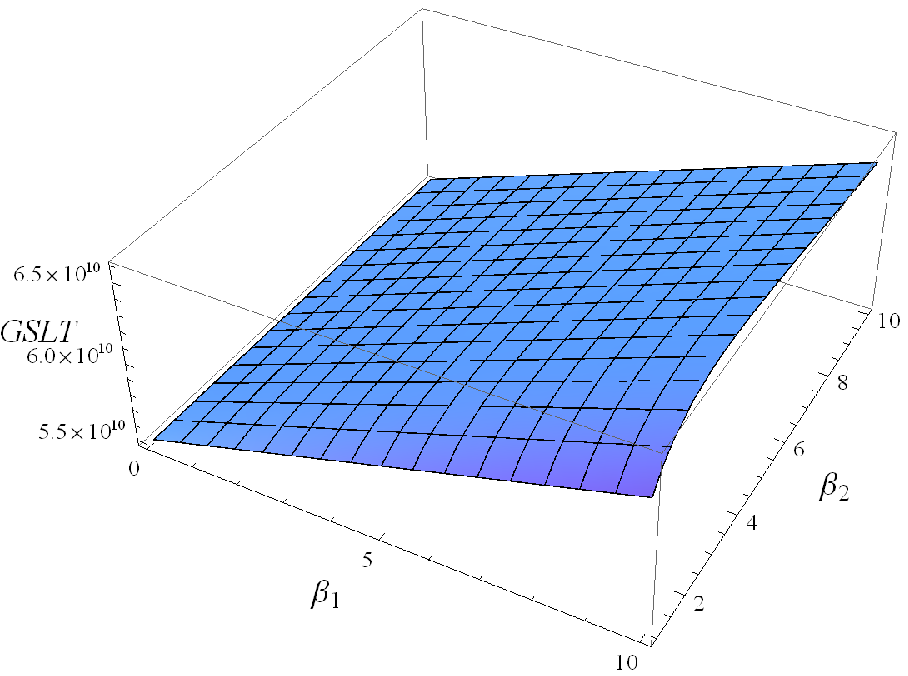, width=0.45\linewidth}
\epsfig{file=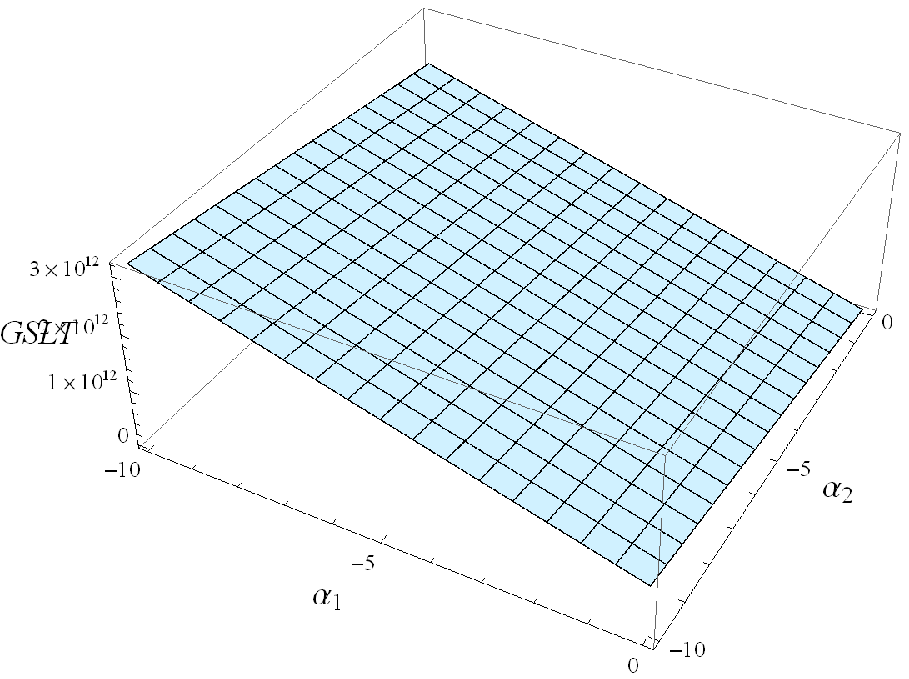, width=0.45\linewidth} \caption{Evolution of
GSLT for the model $f(T,T_G)=-T+\beta_1\sqrt{T^2+\beta_2T_G}+\alpha_1T^2+\alpha_2T\sqrt{|T_G|}$
versus the parameters $\alpha_i (i=1,2)$ and $\beta_i (i=1,2)$.
The left plot corresponds to parameters $\alpha_1=-0.2$ and
$\alpha_2=-0.1$ and right plot corresponds to $\beta_1=0.2$ and
$\beta_2=2$.}
\end{figure}
\begin{figure}
\center\epsfig{file=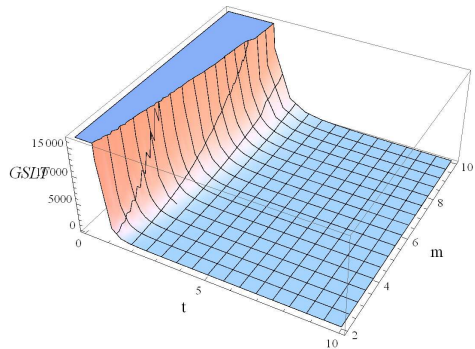, width=0.45\linewidth}
\caption{Evolution of GSLT for the model $f(T,T_G)=-T+\beta_1\sqrt{T^2+\beta_2T_G}+\alpha_1T^2+\alpha_2T\sqrt{|T_G|}$
versus $m$ and $t$ with $\alpha_1=-0.2$, $\alpha_2=0.1$, $\beta_1=0.2$ and $\beta_2=0.1$.}
\end{figure}
\begin{itemize}
  \item $f(T,T_G)=-T+\beta_1(T^2+\beta_2T_G)+\beta_3(T^2+\beta_4T_G)^2$
\end{itemize}
Here, $f(T,T_G)$ model involves fourth order torsion terms and second order
contribution from $T_G$. For this model, the derivatives of $f$ are obtained
as
\begin{eqnarray}\nonumber
f_T&=&-1+2\beta_1T+4\beta_3T(T^2+\beta_4T_G),\\\nonumber
\dot{f}_T&=&(2\beta_1+12\beta_3T^2+4\beta_3\beta_4T_G)\dot{T}+4\beta_3\beta_4
T\dot{T}_G,\\\nonumber
\dot{f}_{T_G}&=&4\beta_3\beta_4T\dot{T}+2\beta_3\beta_4^2\dot{T}_G,\\\nonumber
\ddot{f}_{T_G}&=&4\beta_3\beta_4\dot{T}^2+4\beta_3\beta_4T\ddot{T}+2\beta_3\beta_4^2\ddot{T}_G.
\end{eqnarray}
Using the above expressions we find constraint for GSLT of the form
\begin{eqnarray}\nonumber
T_h\dot{S}_{tot}&=&\frac{-\dot{H}}{2GH^4}(2H^2+\dot{H})-\frac{2\pi}{H^4}(H^2+\dot{H})
\left[4H\{(2\beta_1+12\beta_3T^2+4\beta_3\beta_4T_G)\dot{T}\right.\\\nonumber&+&\left.4\beta_3\beta_4T\dot{T}_G\}
+4\dot{H}\{-1+2\beta_1T+4\beta_3(T^2+\beta_4T_G)T\}+(8H^3-16H\dot{H})\right.\\\nonumber&\times&\left.\{(2\beta_1+12\beta_3T^2+4\beta_3\beta_4T_G)\dot{T}
+4\beta_3\beta_4T\dot{T}_G\}-8H^2\{4\beta_3\beta_4(\dot{T}^2+T\ddot{T})\right.\\\label{37}&+&\left.2\beta_3\beta_4^2\ddot{T}_G\}\right].
\end{eqnarray}
One can see that constraint (\ref{37}) depends only on the parameters
$\beta_1$, $\beta_3$ and $\beta_4$. To specify the values of these parameters
we consider recent cosmic parameters (\ref{34}) and show the evolution of
GSLT in left plot of Figure 5. In right plot we consider the power law
cosmology and fix $\beta_i$ to find variation to find variation of GSLT
versus $m$ and $t$.
\begin{figure}
\center\epsfig{file=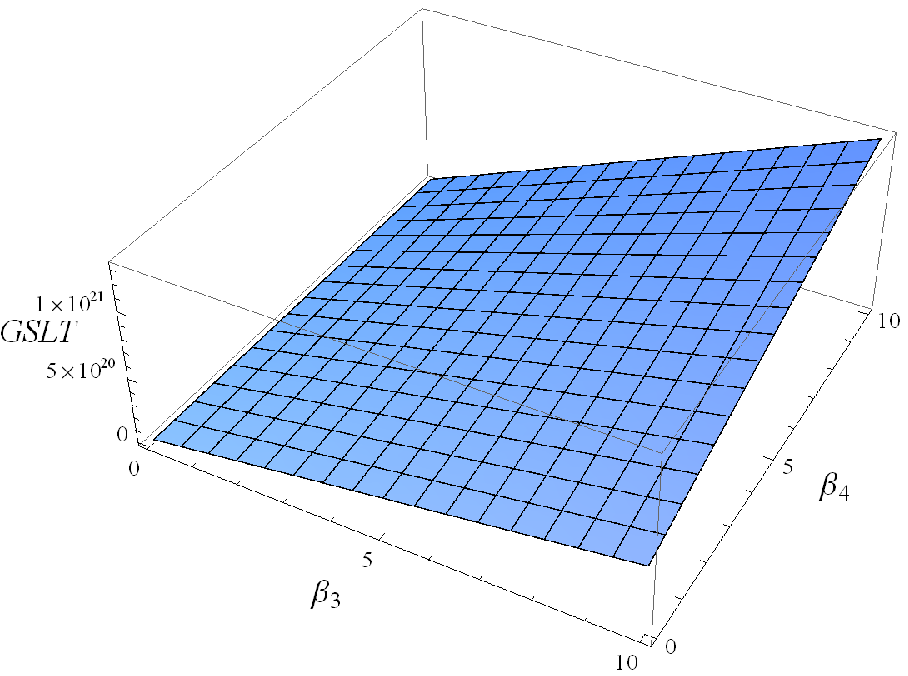, width=0.45\linewidth}
\epsfig{file=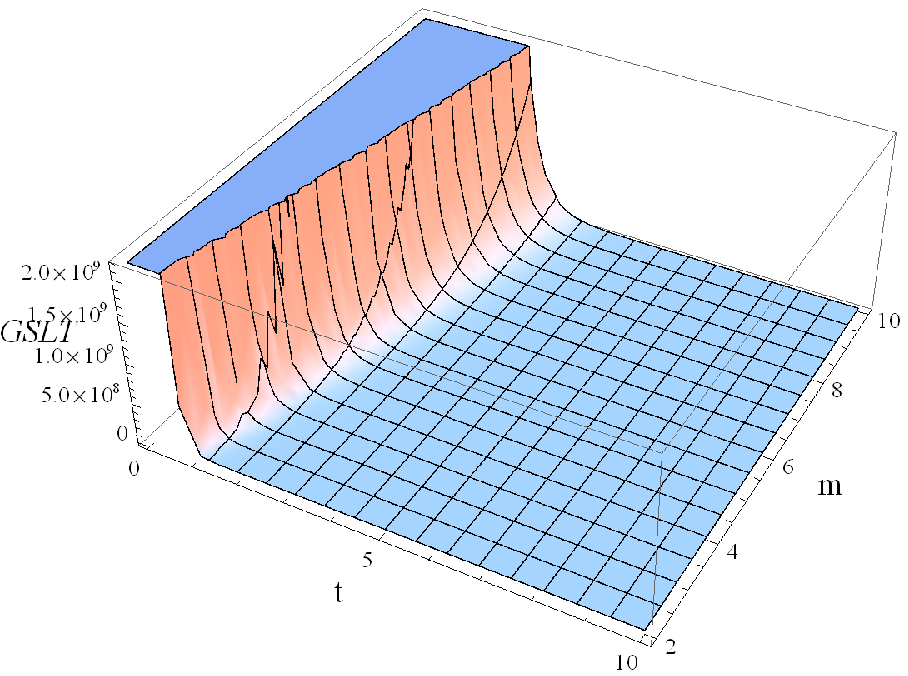, width=0.45\linewidth} \caption{Evolution of
GSLT for the model $f(T,T_G)=-T+\beta_1(T^2+\beta_2T_G)+\beta_3(T^2+\beta_4T_G)^2$.
In left plot we fix the $\beta_1=2$ and show the variation of $\beta_3$ and $\beta_4$. The
right plot shows the evolution of GSLT in terms of parameter $m$ and time $t$ with $\beta_1=.1$,
$\beta_3=1$ and $\beta_4=2$.}
\end{figure}

\section{Concluding Remarks}

In this paper, the thermodynamics properties have been discussed in
$f(T,T_G)$ theory, where $T$ stands for torsion and $T_G$ represents the
teleparallel equivalent of the Gauss-Bonnet term. We present the equilibrium
picture of thermodynamics at the apparent horizon of FRW spacetime. We show
that field equations can be cast to the form of FLT $TdS=-dE+dW$. We find
that no entropy production term is produced in this work as compared to
modified theories involving curvature matter coupling \cite{28}-\cite{30}.
The results of this theory coincide with that in Einstein, Gauss-Bonnet,
Lovelock and braneworld modified theories \cite{8}-\cite{10}.

We also explore the validity of GSLT in the framework of $f(T,T_G)$ gravity.
In this perspective, we consider three generic $f(T,T_G)$ models namely,
$f(T,T_G)=-T+\alpha_1\sqrt{T^2+\alpha_2{T_G}}$,
$f(T,T_G)=-T+\beta_1\sqrt{T^2+\beta_2T_G}+\alpha_1T^2+\alpha_2T\sqrt{|T_G|}$,
and $f(T,T_G)=-T+\beta_1(T^2+\beta_2T_G)+\beta_3(T^2+\beta_4T_G)^2$. We set
the constraint for GSLT in terms of present day values of Hubble,
deceleration, jerk and snap parameters. In Figure 1, we show the evolution of
GSLT for model \textbf{1} and it is found to be satisfied if $\alpha_i>0$. In
this discussion we further consider the power law cosmology and found the
constraints for accelerated cosmic expansion. Figure 2 shows that GSLT can be
met for $m>2$ with $\alpha_1=14$ and $\alpha_2=0.02$. For second model one
requires coupling parameters $\alpha_i<0$ with $\beta_i>0$. Moreover in case
of model \textbf{3}, we fix $\beta_1=2$ and show the variation of $\beta_3$
and $\beta_4$ in left plot of Figure 5. In right plot the evolution of GSLT
for power law cosmology with fixed parameters $\beta_i$.


\vspace{0.25cm}

\begin{thebibliography}{44}

\bibitem{1}Hawking, S.W.: Commun. Math. Phys. \textbf{43}(1975)199;
Bekenstein, J.D.: Phys. Rev. D \textbf{7}(1973)2333.

\bibitem{2}Bardeen, J.M., Carter, B. and Hawking, S.W.: Commun. Math. Phys.
\textbf{31}(1973)161.

\bibitem{3}Jacobson, T.: Phys. Rev. Lett. \textbf{75}(1995)1260.

\bibitem{4}Frolov, A.V. and Kofman, L.: JCAP \textbf{05}(2003)009.

\bibitem{5}Padmanabhan, T.: Class. Quantum Grav. \textbf{19}(2002)5387.

\bibitem{6}Paranjape, A., Sarkar, S. and Padmanabhan, T.:Phys. Rev. D
\textbf{74}(2006)104015; Kothawala, D. and Padmanabhan, T.: Phys.
Rev. D \textbf{79}(2009)104020.

\bibitem{7}Cai, R.G. and Kim, S.P.: JHEP \textbf{02}(2005)050.

\bibitem{8}Cai, R.G. and Cao, L.M.: Nucl. Phys. B \textbf{785}(2007)135.

\bibitem{9}Sheykhi, A., Wang, B. and Cai, R.G.: Nucl. Phys. B
    \textbf{779}(2007)1.

\bibitem{10}Sheykhi, A., Wang, B. and Cai, R.G.: Phys. Rev. D
    \textbf{76}(2007)023515.

\bibitem{11}Akbar, M. and Cai, R.G.: Phys. Rev. D \textbf{75}(2007)084003.

\bibitem{12}Eling, C., Guedens, R. and Jacobson, T.: Phys. Rev. Lett.
\textbf{86}(2006)121301.

\bibitem{13}Cai, R.G. and Cao, L.M.: Phys. Rev. D \textbf{75}(2007)064008.

\bibitem{14}Bamba, K. and Geng, C.Q.: Phys. Lett. B \textbf{679}(2009)282;
    JCAP \textbf{06}(2010)014; ibid. \textbf{11}(2011)008; Bamba, K., Geng,
    C.Q. and Tsujikawa, S.: Phys. Lett. B \textbf{668}(2010)101; Bamba, K.,
Geng, C.Q., Nojiri, S., and Odinstov, S.D.: Europhys. Lett.
\textbf{89}(2010)50003; Bamba, K., Jamil, M., Momeni, D. and Myrzakulov, R.:
Astrophys. Space Sci. \textbf{344}(2013)259.

\bibitem{15}Wu, S.-F., Wang, B., Yang, G.-H. and Zhang, P.-M.: Class.
    Quantum Grav.\textbf{25}(2008)235018;
    Sheylhi, A., Teimoori, Z. and Wang, B.: Phys. Lett. B \textbf{718}(2013)1203.

\bibitem{16}Jamil, M., Saridakis, E.N. and Setare, M.R.: JCAP
    \textbf{11}(2010)032.

\bibitem{17}Wu, S.-F., Wang, B., Yang, G.-H. and Zhang, P.-M.: Class. Quantum
    Grav. \textbf{25}(2008)235018.

\bibitem{18}Sadjadi, H.M.: Phys. Rev. D \textbf{73}(2006)063525; ibid.
    \textbf{76}(2007)104024; Phys. Lett. B \textbf{645}(2007)108.

\bibitem{19}Karami, K. and Abdolmaleki, A.: JCAP \textbf{04}(2012)007.

\bibitem{20}Sheykhi, A., Wang, B. and Cai, R.G.: Nucl. Phys. B
    \textbf{779}(2007)1; Phys. Rev. D \textbf{76}(2007)023515; Sheykhi,
    A.: Phys. Rev. D \textbf{87}(2013)024022.

\bibitem{21}Nojiri, S. and Odintsov, S.D.: Int. J. Geom.Meth. Mod. Phys.
    \textbf{4}(2007)115; Phys.Rept. \textbf{505}(2011)59; De Felice, A., Tsujikawa, S.: Living Rev.
    Rel. \textbf{13}(2010)03; Bamba, K. Capozziello, S. Nojiri, S. and Odintsov,
    S.D.: Astrophys. Space Sci. \textbf{345}(2012)155.

\bibitem{22}Bertolami, O., Harko, T., Lobo, F.S.N. and Pa'ramos, J.:
    \emph{The Problems of Modern Cosmology} (Tomsk State Pedagogical
    University Press, 2009); Harko, T., Lobo, F.S.N., Nojiri, S. and Odintsov, S.D.:
    Phys. Rev. D \textbf{84}(2011)024020; Sharif, M. and Zubair, M.: J. Phys. Soc. Jpn.
    \textbf{81}(2012)114005; ibid. \textbf{82}(2013)014002; ibid.
    \textbf{82}(2013)064001; Houndjo, M.J.S.: Int. J. Mod. Phys. D
    \textbf{21}(2012)1250003; Alvarenga, F.G. et al.: Phys. Rev.
    D \textbf{87}(2013)103526; Odinstov, S.D. and Saez-Gomez, D.: Phys. Lett. B
    \textbf{725}(2013)437; Haghani, Z., Harko, T., Lobo, F.S.N., Sepangi, H.R. and Shahidi,
    S.: Phys. Rev. D \textbf{88}(2013)044023; Sharif, M., Zubair, M.: Gen. Relativ. Gravit.
    \textbf{46}, 1723(2014); Sharif, M., Zubair, M.: Astrophys. Space Sci.
    \textbf{349}, 29(2014).

\bibitem{23}Moller, C.: Mat. Fys. Skr. Dan. Vid. Selsk. \textbf{1}(1961)10;
    Pellegrini, C. and Plebanski, J.: Mat. Fys. Skr. Dan. Vid. Selsk.
    \textbf{2}(1963)4.

\bibitem{24}Harko, T., Lobo, F.S.N., Otalora, G. and Saridakis, E.N.: Phys.
    Rev. D \textbf{89}(2014)124036.

\bibitem{25}Zubair, M. and Waheed, S.: Astrophys. Space Sci.
    \textbf{355}(2014)361.

\bibitem{25a}Zubair, M.: \emph{Thermodynamic laws in modified gravity with non-minimal torsion matter coupling} (Submitted)

\bibitem{26}Kofinas, G. and Saridakis, E.N.: Phys. Rev. D
    \textbf{90}(2014)084044; Phys. Rev. D
    \textbf{90}(2014)084045; Kofinas, G., Leon, G. and Saridakis, E.N.: Class. Quantum Grav.
    \textbf{31}(2014)175011.

\bibitem{27}Waheed, S. and Zubair, M.: arXiv:1503.07413

\bibitem{28}Sharif, M. and Zubair, M.: JCAP \textbf{03}(2012)028; J. Exp.
    Theor. Phys. \textbf{117}(2013)248.

\bibitem{29}Sharif, M. and Zubair, M.: Adv. High Energy Phys.
    \textbf{2013}(2013)947898.

\bibitem{30}Sharif, M. and Zubair, M.: JCAP \textbf{11}(2013)042.

\bibitem{31}Hayward, S.A.: Class. Quantum Grav. \textbf{15}(1998)3147;
Hayward, S.A., Mukohyama, S. and Ashworth, M.: Phys. Lett. A
\textbf{256}(1999)347.

\bibitem{32}Izquierdo, G. and Pavon, D.: Phys. Lett. B \textbf{633}(2006)420.

\bibitem{33}Rapetti, D. et al.: Mon. Not. R. Astron. Soc.
    \textbf{375}(2007)1510; Riess, A. G. et al.: Astrophys. J.
    \textbf{730}(2011)119; N. J. Poplawski: Class. Quantum Grav. \textbf{24}(2007)3013.



\end{thebibliography}
\end{document}